\documentclass{aa}

\usepackage{url}

\usepackage{graphicx}
\usepackage{natbib}
\bibpunct{(}{)}{;}{a}{}{,} 
\usepackage{amssymb}

\newcommand{\eqn}[1]{\begin{center}\begin{equation} #1 \end{equation}\end{center}}

\newcommand{\Hp}{\mathrm{H}_{\mbox{p}}}
\newcommand{\Msol}{\mathrm{M}_{\sun}}

\newcommand{\dex}{dex}

\newcommand{\teff}{\mathrm{T_{eff}}}

\newcommand{\basel} {{\it B}a{\it S}e{\it L}}

\def\lg{$\log g$}

\begin{document}
\title{Convective Core Mixing: a Metallicity Dependence?}

\author{     D. Cordier\inst{1,2}
        \and Y. Lebreton\inst{1}
        \and M.-J. Goupil\inst{1}
        \and T. Lejeune\inst{4}
        \and J.-P. Beaulieu\inst{5}
        \and F. Arenou\inst{1}
       }

\institute{DASGAL, CNRS UMR 8632,
           Observatoire de Paris-Meudon, DASGAL, 
           F-92195 Meudon Principal Cedex, France.
   \and    \'Ecole Nationale Sup\'erieure de Chimie de Rennes, 
           Campus de Beaulieu, F-35700 Rennes, France
   \and    Observatorio Astronomico, Universidade de Coimbra,
           Santa Clara 3040 Coimbra, Portugal.
   \and    I.A.P., 98bis boulevard Arago, F-75014 Paris, France.}
           
\offprints{daniel.cordier@ensc-rennes.fr}
  
\date{Received ; accepted}

\markboth{Cordier et al., Core Mixing and Metallicity Dependence}{}

\abstract{
The main scope of this paper is to investigate  the possible existence of a
metallicity dependence of the overshooting from main sequence stars turbulent cores.
We focus on objects with masses in the range $\sim 2.5 \Msol - 
\sim 25 \Msol$. Basically, evolutionary time scale ratios are compared  with star numbers
ratios on the main sequence. Star populations are synthesized using grids of evolutionary
tracks computed with various overshooting amounts. Observational material is provided by
the large and homogeneous photometric database of OGLE 2 project for the Magellanic clouds.
Attention is paid to the study of uncertainties: distance modulus, intergalactic and interstellar
reddening, IMF slope and average binarity rate. Rotation and chemical composition
gradient are also considered. The result for the overshooting distance is 
$l_{\mathrm{over}}^{\mathrm{ SMC}}= 0.40^{+0.12}_{-0.06} \mathrm{H}_{\mathrm{p}}$ 
($Z_{0}=0.004$) and
$l_{\mathrm{over}}^{\mathrm{ LMC}}= 0.10^{+0.17}_{-0.10} \mathrm{H}_{\mathrm{p}}$
($Z_{0}=0.008$) suggesting a possible  dependence of the extent of
the mixed central regions  with metallicity within the considered mass range. Unfortunately it is
not yet possible to fully disentangle effects of mass and chemical composition. 
}

\maketitle

\section{\label{intro}Introduction}

Large convective phenomena  occur in the cores of main sequence stars with masses 
above about 1.2 $\Msol$ (for galactic chemical composition).  
In standard models, convection is crudely modeled with the  well-known Mixing Length 
Theory of \citet{bohm_vitense58} (hereafter MLT) and the core extension is determined 
according to the Schwarzschild criterion. The Schwarzschild limit is the value of the 
radius where the buoyancy force vanishes. However  inertia of the convective elements 
leads to an extra mixing above the Schwarzschild limit, called ``overshooting'' 
and usually expressed in fraction of the pressure scale height. Several theoretical works 
\citep[for a review see][]{zahn_91} give arguments in favor of such an additional mixing. 
Many laboratory experiments show evidences for overshooting \citep[see][]{Massaguer90}. 
Although overshooting can occur below an external convective zone \citep[see][]{alongi_etal_91}, 
this paper is  exclusively concerned with core overshooting.

 One of the first empirical determination of 
convective core overshooting was obtained by \citet{Maed:Mermi}
who used a set of 34 galactic 
open clusters and fitted the main sequence width with an additional 
mixing of about 20-40 \% in mass fraction. \citet{mermilliod_86} derived an overshooting amount of
about $0.3 \; \mathrm{H}_{\mathrm{p}}$ again  for solar-like chemical composition and for a 9-15 $\mathrm{M}_{\sun}$
range. \citet{stothers_chin_91a} derived an overshooting amount $< 0.2 \; \mathrm{H}_{\mathrm{p}}$
for Pop. I stars using the metal-enriched opacity tables published in \citet{rogers_iglesias_92}.

During the last decade, many evolutionary model grids have been computed with an
overshooting amount equal or close to $0.2 \; \mathrm{H}_{\mathrm{p}}$: e.g.
\citet{charbonnel_etal_96} or \citet{bertelli_et_al_94}. This second team uses
a formalism \citep[see][]{bressan_etal_81} slightly different from the Geneva
team one. Generally the same overshooting amount is used whatever the
metallicity and mass are.

 \citet{Kozhurina-Platais_etal_97} obtained $l_{\mathrm{over}}= 0.2 \pm 0.05 \; \mathrm{H_{P}}$
for the galactic cluster NGC 3680 (solar metallicity) with the 
isochrone technique, this method consists in fitting the cluster CMD features (particularly
the turn-off position) with model isochrones. \citet{Iwamo99} compared evolutionary models with observations 
of three binary systems: V2291 Oph, $\alpha$ Aur and $\eta$ And (``binary system'' technique). The authors
adjusted either the helium content or the overshooting parameter to get a better fit to observations.
The best results were obtained  with  a moderate overshooting amount ($\la 0.15 \; \Hp$). For
super-solar metallicity ($Z_{0}= 0.024$) \citet{lebreton_etal_2001} derived $l_{\mathrm{over}} \lesssim 0.2
\; \mathrm{H_{P}}$ from the modeling of the Hyades cluster turn-off.

\citet{Maed:Mermi} have suggested an overshooting increasing with mass within the studied
range of 2-6 $\mathrm{M}_{\sun}$ which is also found
 by \citet{schroder_etal_97} with a 
study of binary systems. According to their results,
 the overshooting should increase from
$\la 0.24 \;\Hp$ for 2.5 $\mathrm{M}_{\sun}$ to $\la 0.32 \; 
\Hp$ for 6.5 $\mathrm{M}_{\sun}$. With a similar study
\citet{ribas_etal_00} also found a mass dependence.

 The question of a metallicity dependence must also be addressed. 
\citet{ribas_etal_00}'s results suggest  
a slight metallicity dependence
for a stellar mass around 2.40 $\mathrm{M}_{\sun}$ (see their Table 1). 
The more metal poor star SZ Cen 
(in mass fraction: $Z_{0}=0.007$) is satisfactorily modelled with 
 an overshooting distance  $0.1 \;\mathrm{H}_{\mathrm{p}}\lesssim l_{\mathrm{ov}} 
\lesssim 0.2 \; \mathrm{H_{P}}$ and objects 
with $Z_{0}$ ranging between 0.015 and 0.020
seem to have an overshooting around 0.2 $\mathrm{H_{P}}$.
 \citet{keller_etal_2001} have recently explored 
the dependence of overshooting with
metallicity by means of the isochrone technique using isochrone grids 
from the Padova group. Their study involves  HST observations 
of four clusters: NGC 330 (SMC), 
1818, 2004 and 2001 (LMC). \citet{keller_etal_2001}
 find the best fit (with respect to age and overshooting) for 
an overshooting amount which is equivalent to 
$l_{\mathrm{over}}= 0.31 \pm 0.11 \; \mathrm{H}_{\mathrm{p}}$ in the Geneva formalism
($l^{\mathrm{Padova}}_{\mathrm{over}}= 2 \times l^{\mathrm{Geneva}}_{\mathrm{over}}$).
 
 In this paper, we carry out an independent study of a possible metallicity
 dependence of overshooting with a  technique which differs from 
the ``binary system''  \citep{ribas_etal_00,andersen_91}
and ``isochrone'' techniques. Our 
method is based on star-count ratios, with comparisons between
observational material and synthetic population results in color-magnitude
 (CMD) diagrams. We are then led to discuss several points:
particularly distance modulus, reddening and binarity rate.
 If the dependence of overshooting with metallicity (or mass) was thereby to be firmly
assessed, it would then be a challenge to understand its
physical origin. 

   \begin{figure*}
   \centering
\rotatebox{-90}{\resizebox{11cm}{18cm}{\includegraphics{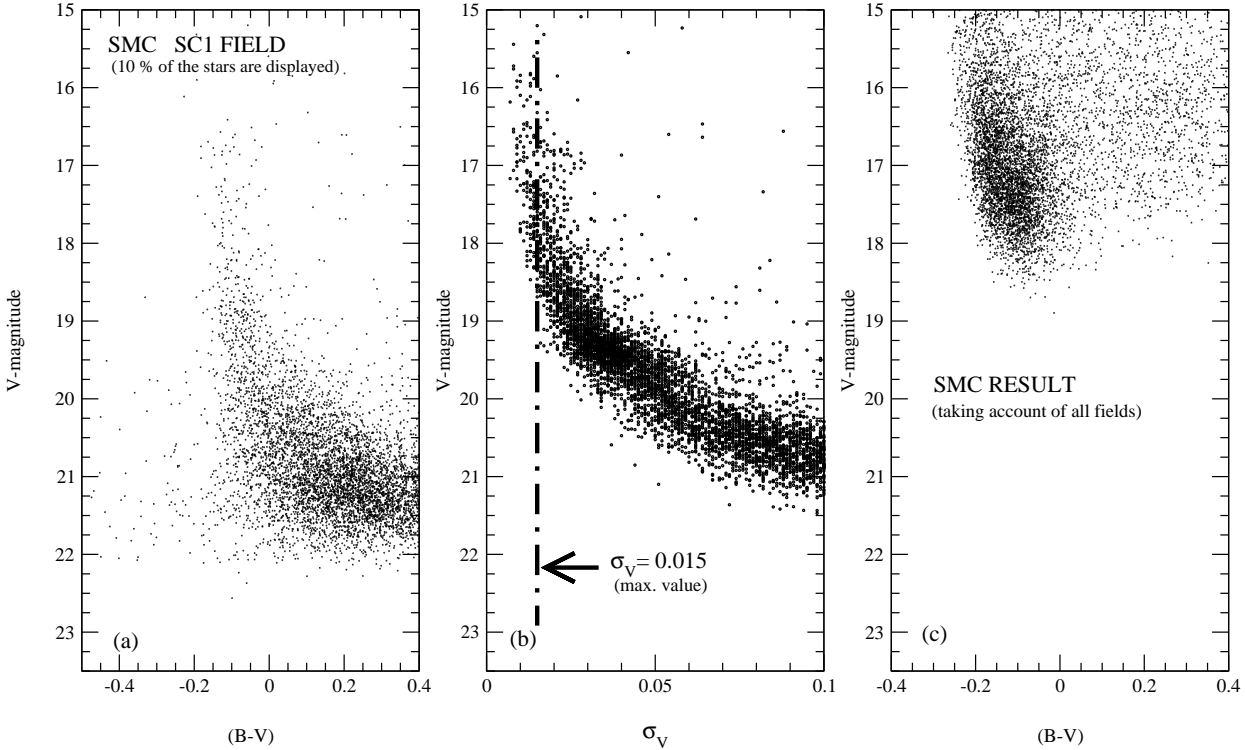}}}
      \caption{(a) CM-Diagram for the OGLE 2 SC 1 field, 10 \% of the data have been plotted
               for sake of clarity. (b) Standard deviation of the measurements: 
               $\sigma_{V}$ versus magnitude $V$, the dot-dashed line indicates the
               limit-value considered (0.015). (c) The resulting CM-diagram after selection  
               (all the fields within SMC have been plotted).}
         \label{data_select_SMC}
   \end{figure*} 

We are concerned with  a metallicity range relevant to the Magellanic Clouds and  take
advantage of the homogeneous OGLE 2 data, which provide color magnitude diagrams for 
$\sim 2 \times 10^{6}$ stars in the Small Magellanic Cloud  (hereafter SMC) and 
$\sim 7 \times 10^{6}$ in the Large Magellanic Cloud (hereafter LMC).
On the theoretical side, we estimate the number of stars from evolutionary
model sequences computed with different amounts of overshooting. 
 From these data 
sets and using evolutionary models with intermediate and low metallicity, 
we estimate the overshooting value during the main sequence in the SMC and LMC
for a stellar mass in the range 2.5 $\mathrm{M}_{\sun}$ - 25 $\mathrm{M}_{\sun}$.

 In Sect.~\ref{data} we describe the observational data involved in this work. 
Sect.~\ref{method} is devoted to the
method used: data selection and star counting. 
Sect.~\ref{popsynth} gives the main features of our population
synthesis procedure. 
Sect.~\ref{astro_in} is devoted to astrophysical inputs, and Sect.~\ref{results} to results and 
effects of uncertainties.  Sect.~\ref{discu} discusses the results. 
 It must be emphasized that we determine  in fact the extent of the inner
mixed core region which can be due either to true overshooting or to another
process as rotation, some observational evidences exist about correlation betweenn metallicity and
$v \sin i$ (see Venn et al. 1999), the problem of rotation is shortly discussed in the Sect.~\ref{discu}.
Finally, Sect.~\ref{concl} gives some 
comments and concluding remarks. An appendix has been added to provide details
 about the 
population synthesis algorithm and error simulations.

\section{\label{data}Observational data}

 The observational data set considered here has been obtained  by the
Optical Gravitational Lensing Experiment (OGLE hereafter) consortium during its second 
operating phase (for more details and references the reader can consult URL: 
\url{http://www.astrouw.edu.pl/~ogle/}).

\subsection{SMC and LMC data}

 We have downloaded the SMC data described in \citet{udalski98}.
The data used in this paper are from the post-Apr. 8, 2000 revision. The SMC is divided into 11 fields
(labeled SC1 to SC11) covering 55'x14'; each field contains between $\sim 100,000$ and 
$\sim 350,000$ objects. For each object several quantities are available: equatorial coordinates,
BVI photometry and associated standard errors $\sigma_{B}$, $\sigma_{V}$ and $\sigma_{I}$.
This database has the great advantage of being extensive and very homogeneous.

 The LMC data are described in \citet{udalski00}. The BVI map of the LMC is composed of 26 
fields (SC1 to SC26) in the central bar of the LMC. The dataset includes photometry and astrometry
for about 7 million stars over a 5.7 square degree field.
   \begin{figure}
   \centering
\rotatebox{-90}{\resizebox{6.1cm}{8.8cm}{\includegraphics{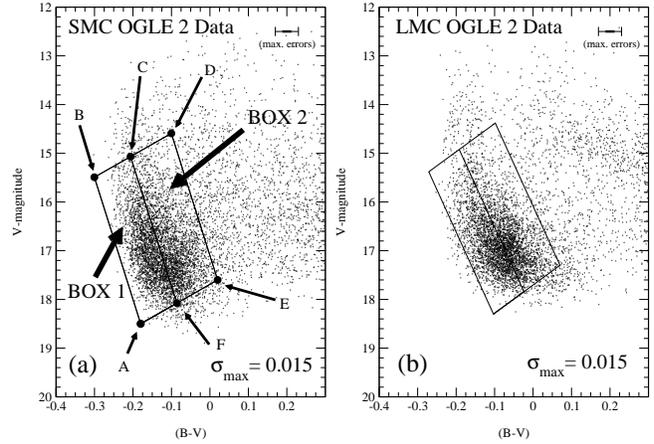}}}
      \caption{(a) Data from SMC with $\sigma_{V} \le 0.015$ mag and box definitions
               ($N_{1}+N_{2}= 4653$ and $N_{2}/N_{1}= 1.08$), 
               (b) the same for the LMC case ($N_{1}+N_{2}= 4113$ and $N_{2}/N_{1}= 1.01$).}
         \label{smc_lmc_boxes}
   \end{figure}
\section{\label{method}The star-count method}

\subsection{\label{dataselec}Data selection}

 As shown in Fig.~\ref{data_select_SMC}b, 
the standard error on V-magnitude, $\sigma_{\mathrm{V}}$,
increases with the magnitude. This is also true  
for B or I-magnitudes. Hence the errors on 
$(\mathrm{B-V})$ or $(\mathrm{V-I})$   colors
rapidly  increase and reach values as large as 0.2 mag around a
V-mag$\sim 20$: this is of the same order as the Main Sequence width.

As we are interested in  the MS structure and 
as we must minimize error effects while keeping a 
quite good statistics, we have chosen to take into 
account only  data with $\sigma_{\mathrm{V}}$ and 
$\sigma_{\mathrm{B}}$ (or $\sigma_{\mathrm{I}}$) 
lower or equal to $0.015$ mag, leading to a maximum error
on color of $0.02$ mag. 
The value of 0.015 mag appears to be an optimal choice 
maintaining a good statistics
with photometric errors remaining small compared with the MS width. 
Fig.~\ref{data_select_SMC} sketches
the proposed selection process and displays differences 
between the entire Color-Magnitude Diagram
(Fig.~\ref{data_select_SMC}a) and 
the final diagram (Fig.~\ref{data_select_SMC}c): obviously, the remaining
data are those corresponding to lower magnitudes. 

 This selection process leaves $\sim 4700$ objects on the SMC MS 
(over a total of more than 2.2 millions objects) in the BV system ($\sim$1100 objects in the VI system)
and $\sim$ 4000 objects on the LMC MS (over a total of more than 7.2 millions objects) in the 
BV system ($\sim$1600 objects in the VI system). As we can see the BV system presents
a more favorable statistics, therefore in the following we will work only with
this set of bands.

 Tables 4 from \citet{udalski98,udalski00} indicate that completeness for 
$\mathrm{V} \lesssim 18$ should be better than about 99 \% for the SMC; and 
should be around 96 \% - 99 \% depending on the field crowding for the LMC.

\subsection{\label{def_boxes}Star count ratios: an observational constraint}

As the absolute number of stars arriving on the ZAMS per
unit of time for a given mass is unknown, 
 we rather compute star count ratios. 
To count stars, we first  define an area in the 
CM-diagram. As we are interested in the MS structure, we choose a region
which contains the main sequence ``bulge'' revealed after the data selection
process (see Fig.~\ref{smc_lmc_boxes}a) with the most convenient  geometrical
shape: a ``parallelogram'' 
(for automatic 
count purpose). A couple of opposite sides 
(AB and DE in Fig.~\ref{smc_lmc_boxes}a) are chosen to be more or less ``parallel'' 
to the main sequence axis.

In the CM-Diagram, main sequence stars evolve from the blue  to the red side. 
The MS width is mainly an evolutionary effect connected 
to a characteristic time scale $\tau_{\mathrm{ MS}}$
(time spent by a star on the Main Sequence). The distribution of the objects within the
Main Sequence should be related to this time scale. Therefore we divide our
parallelogram into two regions called ``box 1'' and ``box 2'' (see Fig.~\ref{smc_lmc_boxes}a)
where the respective numbers of objects $N_{1}$ and $N_{2}$ are similar 
($N_{2}/N_{1} \sim 1$). This ratio is taken as an observational constraint
and it will enable us to discriminate between theoretical grids of evolutionary tracks computed
with various overshooting amounts.

 We now turn to  the method used to build a synthetic stellar sample comparable to the OGLE 2 
ones (after selection) from evolution simulation
outputs.

\section{\label{popsynth}Population synthesis}

\subsection{\label{models}Evolutionary models}

 Our evolutionary models are built with the 1D Henyey type code 
CESAM\footnote{CESAM : Code d'Evolution Stellaire Adaptatif
et Modulaire} \citep[see][]{Mor97} in which we brought several improvements. Applying modern 
techniques like the projection of the solutions on B-spline basis and automatic mesh refinements, 
CESAM allows robust, stable and highly accurate calculations. We use as physical inputs: 
\begin{itemize}
        \item the OPAL 96 opacities from \citet{iglesias96} at high temperatures 
              ($T > 10 000$ K) and  the \citet{alex:fergu94} opacities for cooler 
              domains. For metallicity higher than the solar one (that occurs during 
              the He core burning phase) we use elemental opacities (Los Alamos) 
              calculated by \citet{magee95}.
        \item the EFF equation of state from \citet{Eggleton73}
        \item elemental abundances are from \citet{Grevesse93} (the ``GN93'' 
              mixture), the cosmological helium is from \citet{Izotov97}:
              $Y_{P}= 0.243$, and the helium 
              content is scaled on the solar one following a standard 
              helium-metallicity relation: $Y=Y_{\mathrm{P}}+Z (\Delta Y/\Delta Z)$.
              The calibration of a solar model in luminosity yields 
              $\Delta Y/\Delta Z = 2$ \citep{lebreton99} from the calibration of the solar
              model radius. This value is compatible with the recent value $\Delta Y/\Delta Z= 2.17\pm 0.40$
              of \citet{peimbert00}. We therefore adopt  $\Delta Y/\Delta Z
              \approx 2$.
        \item for the chemical composition we adopt $\mathrm{[Fe/H]}$ derived from Cepheid
              measurements by \citet{luck_et_al98}:
              \begin{itemize}
                \item for the SMC they find a range from $-0.84$ to $-0.65$ with a mean value:
                      $\mathrm{[Fe/H]}= -0.68$ which leads to $X_{0}=0.745$, $Y_{0}=0.251$ and $Z_{0}=0.004$.
                \item for the LMC they find a range from $-0.55$ to $-0.19$, combining all the values we
                      obtain a mean value of $\mathrm{[Fe/H]}= -0.34$ leading to:
                      $X_{0}= 0.733$, $Y_{0}= 0.259$ and $Z_{0}= 0.008$.
              \end{itemize}
        \item The nuclear reaction rates are from \citet{caughlan88}, 
              except:  
              $^{12}\mathrm{C}(\alpha,\gamma)^{16}\mathrm{O}$, 
              $^{17}\mathrm{O}(p,\gamma)^{18}\mathrm{F}$ 
              taken from \citet{caughlan85} and  
              $^{17}\mathrm{O}(p,\alpha)^{14}\mathrm{N}$ taken from \citet{landre90}.
              The adopted rate for \element[][12]{C}($\alpha,\gamma$)\element[][16]{O}
              is quite similar to the NACRE compilation \citep{angulo_etal_99} one: a factor of 
              about two higher than \citet{caughlan88} and about 80\% of
              Caughlan et al.'s (1985) one.
        \item To take into account the metallicity effect on the mass loss rate
              \citep{jager88} we adopt the scaling factor 
              $(Z_{0}/0.02)^{0.5}$ derived from the \citet{kudri86} models.
        \item The convective flux is computed according to the classical MLT.
              We use a mixing length value $l_{\mathrm{MLT}}= 1.6 \; \mathrm{H_{P}}$.
              This value has been derived by \citet{schaller_et_al92} from
              the average location of the red giant branch of more than 75 clusters.
              Very similar value (1.64) has been found more recently by 
              \citet{lebreton99}.
              An extra-mixing zone is added above the Schwarzschild convective core:
              this ``extra-mixing'' zone is set to extend over the distance
              $l_{\mathrm{over}}= \alpha_{\mathrm{over}}\; \mathrm{H_{P}}$, $\alpha_{\mathrm{over}}$
              being a free parameter, the value of which is discussed here.
        \item the external boundary conditions are determined in a layer within
              a simple grey model atmosphere built with  an Eddington's $T(\tau)$ law.
\end{itemize}

\subsection{Conversion of the theoretical quantities into observational ones}

 In order to compare theoretical results to observational data, conversions are needed.
Transformations of the theoretical quantities, ($M_{\mathrm{bol}}$,  $\mathrm{T_{eff}}$)
into absolute magnitudes and colors are derived from the most recent version of the
Basel Stellar Library ({\basel}, version 2.2), available electronically at 
{\tt ftp://tangerine.astro.mat.uc.pt/pub/BaSeL/}. This library 
provides  color-calibrated theoretical
flux distributions for  a large range of  
fundamental stellar parameters, 
$T_{\rm eff}$ (2000 to 50,000 K), $\log g$ (-1.0 to 5.5 \dex), 
and $\mathrm{[Fe/H]}$ (-5.0 to +1.0 \dex).
The {\basel} flux distributions are calibrated on 
the stellar $\mathrm{UBVRIJHKL}$ colors, using:
\begin{itemize}
  \item {\em  empirical}  photometric calibrations for solar metallicity
  \item {\em semi-empirical} relations constructed from the color differences 
        predicted by stellar model atmospheres for non-solar metallicities.
\end{itemize}
 Details about the calibration procedure are given in \citet{lejeune97} and \citet{lejeune98}.  
Compared to the previous  versions of the {\basel}  library, all the model spectra of stars with
{$\teff$} $\geq 10,000$ K are now calibrated on empirical colors from the $\mathrm{T_{eff}}$ {\it
versus} $\mathrm{(B-V)}$ relation of \citet{flower96}. In addition, the calibration procedure for 
the cool giant model spectra  has been extended in the present models to the parameter ranges 2500 K 
$\leq$ {$\teff$} $<$ 6000 K and -1.0 $\leq$ {\lg} $<$ 3.5
\footnote{In the  previous  versions   of the {\basel} models, we adopted {$\teff$} = 5000 K and {\lg} = 2.5 
as the upper limits  for  the calibration of giants \citep[see][]{lejeune98}.}.
\subsection{Population synthesis}

 In contrast with ``classical'' works on population synthesis where the
CMD \textbf{as a whole} is simulated, we construct a small part
of the CMD: the area containing the brighter MS stars. In this way the task is simplified.
Artificial stellar samples have been generated from our evolutionary tracks with a specially
designed population synthesis code \verb+CReSyPS+\footnote{Code Rennais de Synth\`ese de Populations Stellaires}.

 In our framework the main hypothesis is that the Star Formation Rate (SFR) is constant
during the time scales involved here: i.e. a few hundred megayears. So for a given mass the
number of observed stars (i.e. those corresponding to a given evolutionary track) must be
proportional to the time scale of the main sequence. We assume that the SFR is constant in time
and mass (equal for all masses in the range explored in this work), if we note $r$ the SFR:
$\Delta t \approx 1/r$ represents the mean time elapsed between two consecutive star births.
For the observational star samples, $\Delta t$ is unknown but the objects numbers are
available. We choose $\Delta t$ to get similar total star numbers in boxes 1 and 2 (i.e. $N_{1}+N_{2}$)
both in the synthetic CMD and observational diagram. We point out that the ratios $N_{2}/N_{1}$ are
not sensitive to the $\Delta t$ value chosen.

The evolutionary track grids scan a mass range between 2.5 $\Msol$ and 25 $\Msol$ from the ZAMS to 
$\log \teff \sim 3.8$ covering the entire 
box ranges in color and magnitude (defined in Sect.~\ref{def_boxes}) .
The mass step is  increasing from 0.5 $\Msol$ around 3 $\Msol$ stars to
5 $\Msol$ above 15 $\Msol$. 
Several overshooting amounts have been used from 0.0 
to 0.8. \verb+CReSyPS+ treats the photometric errors by simulating OGLE 2 ones 
(see App.~\ref{appendix_noise}) which is very important for our purpose.
   \begin{figure*}
   \centering
\rotatebox{0}{\resizebox{18cm}{18cm}{\includegraphics{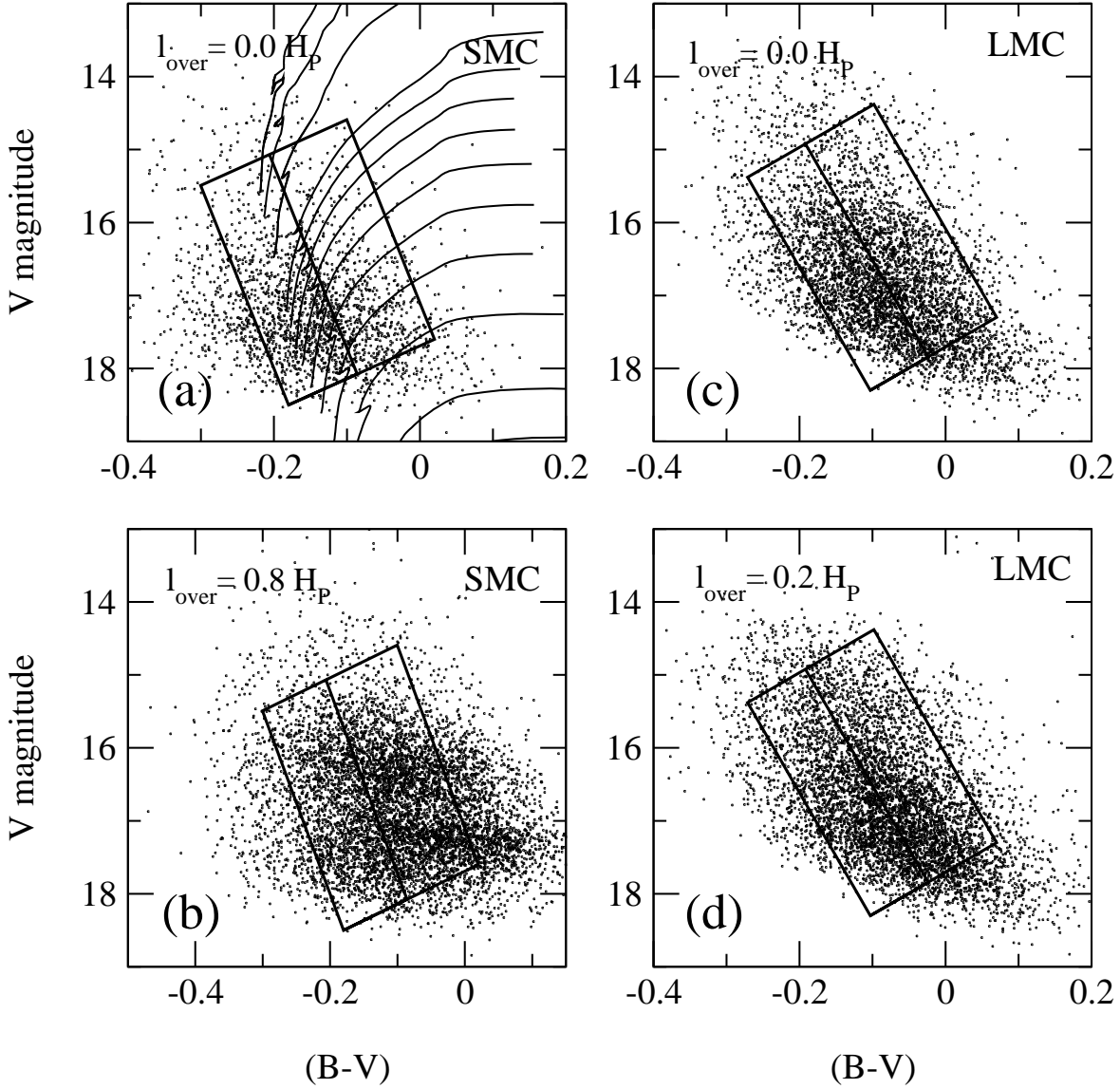}}}
      \caption{Synthetic CM-Diagrams for SMC and LMC chemical compositions, panels (a) and (b) are for
               the SMC with two overshooting amounts: $\alpha_{over}=0.0$ and $\alpha_{over}=0.8$, panels
               (c) and (d) are for the LMC with: $\alpha_{over}=0.0$ and $\alpha_{over}=0.2$ respectively.
               For panel (a) only 30\% of the synthetic objects have been displayed for clarity
               purpose and corresponding evolutionary tracks have been plotted. For other panels:
               (b), (c) and (d) the number of displayed objects has not be reduced.
               In all cases the total number of stars in the boxes -used for calculations- is 
               close to the observational one:
               (a) $N_{1}+N_{2}= 4403$ ($N_{2}/N_{1}= 0.65$), (b) $N_{1}+N_{2}= 5011$ ($N_{2}/N_{1}= 1.59$), 
               (c) $N_{1}+N_{2}= 3582$ ($N_{2}/N_{1}= 0.94$), (d) $N_{1}+N_{2}= 4717$ ($N_{2}/N_{1}= 1.13$).
               We recall that \textbf{empirically} we got for the SMC $N_{1}+N_{2}= 4653$ ($N_{2}/N_{1}= 1.08$) and
               for the LMC $N_{1}+N_{2}= 4113$ ($N_{2}/N_{1}= 1.01$). The ratios $N_{2}/N_{1}$
               obtained theoretically are rather independant from the value of $N_{1}+N_{2}$ in the synthetic
               CM-Diagram, for instance in the case of the panel (b) we got $N_{2}/N_{1}= 1.60$ with
               $N_{1}+N_{2}= 6399$. The cloud of dots in panel (b) ($l_{over}=0.8 \; \Hp$, SMC) calls a
               comment: it appears to be bimodal, i.e. showing over-populated regions around $V \sim 17.5$ mag
               and $V \sim 16.3$ mag. Indeed for high overshooting values the main sequence of masses as small
               as $\sim 3 \Msol$ can reach $V \sim 17.5$ generating with their large evolutionary time scale an
               over-populated region. Moreover the binarity shifts a part of this population to a $\sim 0.8$ mag
               brighter region, with a binarity rate $\beta= 0.0$, this ``bimodal effect'' disappears
               }
         \label{synth_CMD}
   \end{figure*} 
 Our algorithm requires the knowledge of some input parameters: 
distance modulus, reddening 
and absorption, binarity rate, Initial Mass Function (hereafter IMF) slope and photometric errors.

We summarize here the main steps of the algorithm:
\begin{itemize}
  \item {\bf STEP 1:} a mass distribution is generated between 2.5 $\Msol$ and 25 $\Msol$ following the
        Salpeter's law: $dN/dm \approx m^{-\alpha_{\mathrm{Salp}}}$ (see Sect.~\ref{imf_slope}).
  \item {\bf STEP 2:} for each mass, an evolutionary track is interpolated within the grid calculated
        by the evolutionary code. On each track, models are selected every time step $\Delta t$, which is
        adjusted in order to yield  a total number of stars equivalent to the observed one.
  \item {\bf STEP 3:} consistently with the value of the binary rate $<\beta>$
  (see Sect.~\ref{bin_rate}),  objects are randomly selected  to belong to a binary system 
         and the magnitudes of these systems are calculated. Triple systems (and higher 
        multiplicity systems) are neglected.
  \item {\bf STEP 4:} distance modulus is added (and in the case of SMC a random ``depth''
        inside the cloud) and  synthetic photometric errors are attributed to magnitudes 
        (see App.~\ref{appendix_noise}).
  \item {\bf STEP 5:} we use a ``quality filter'': objects with  too large photometric errors are
        rejected from the synthetic sample.
  \item {\bf STEP 6:} color is calculated, reddening and extinction coefficient are applied. Concerning
        reddening, a gaussian distribution is applied around the mean value in order to simulate
        object-to-object variations (see discussion in Sect.~\ref{red_abs}).
\end{itemize}

 With the interpolation between evolutionary tracks,
 it is very important (particularly
at low mass, i.e. $3.0 \Msol \lesssim M \lesssim 4.0 \Msol$) to reproduce the time
scale $\tau_{\mathrm{MS}}$ with a good accuracy. A test at 3.25 $\Msol$ has shown that the
``interpolated time scale'', $\tau^{\mathrm{interpol}}_{\mathrm{MS}}$, is very close to the
calculated one (with the evolutionary code) $\tau^{\mathrm{cal}}_{\mathrm{MS}}$ with a
difference not larger than about $1 \%$. Also important are the magnitude
interpolations on the Main Sequence: our tests also show 
 a very good agreement between interpolated
magnitudes and calculated ones, differences are unsignificant (about $10^{-3} - 10^{-2}$ mag,
whereas the photometric errors are much larger).

 Our code intensively uses a random number generator. 
We have chosen an algorithm insuring a very large
period about $2 \times 10^{18}$ 
\citep[program ``\textbf{ran2}'' from][]{Num_Recipes},
which is much larger than  the number of synthetized objects.

 As a result, examples of synthetic samples generated by \verb+CReSyPS+ are displayed
in Fig.~\ref{synth_CMD} where the influence of overshooting is shown for both clouds.

\section{\label{astro_in}Astrophysical inputs}

\subsection{Distance modulus}
\paragraph{Large Magellanic Cloud.}
 The LMC distance modulus has a key role in extragalactical distance determinations, but its value
is still debated. The determinations range between ``short'' distance scales
\citep[i.e.][]{stanek_etal_98} and ``long'' distance scales \citep[i.e.][]{laney_stobies94}. Using
the HIPPARCOS calibrated red clump stars, \citet{stanek_etal_98} found 
$\mu_{0,\mathrm{LMC}} = 18.065 \pm 0.031 \pm 0.09$ mag and \citet{laney_stobies94} from a study
of Cepheids Period-Luminosity relation obtained $\mu_{0,\mathrm{LMC}} = 18.53$ mag with an internal error of 
$0.04$ mag.
 \citet{groenewegen_salaris_01} found $\mu_{0,\mathrm{HV 2274}} =18.46 \pm 0.06$ mag from a study of 
the LMC-eclipsing binary system HV 2274. They indicate a LMC center distance at
$\mu_{0,\mathrm{LMC}}= 18.42 \pm 0.07$ mag. Recently, from the DENIS survey data, \citet{cioni_etal_00}
derived a distance modulus for the LMC of $\mu_{0,\mathrm{LMC}}= 18.55 \pm 0.04$ (formal) $\pm 0.08$ 
(systematic) using a method based on the apparent magnitude of the tip of the red giant branch.
 The HST Key Project Team adopted $\mu_{0,\mathrm{LMC}}= 18.50 \pm 0.15$ mag \citep{mould_etal_00}.
In order to bracket the most recent estimations,
 we have chosen the HST Key Project value:
$$
\mu_{\mathrm{LMC}}= 18.50 \pm 0.15 \;\mathrm{mag}
$$
 \citet{van_der_Marel_etal_01} give an order of magnitude of the depth of the
LMC. They indicate small corrections to magnitude 
for well studied individual objects within the LMC,
ranging  between $\Delta\mu_{0,\mathrm{LMC}}= -0.013$ (SN 1987A) to
$\Delta\mu_{0,\mathrm{LMC}}= +0.015$ (HV 2274). We neglect these corrections which have the same order 
of magnitude than the photometric errors.
\paragraph{Small Magellanic Cloud.}
 \citet{laney_stobies94} suggest a distance modulus (based on Cepheids) of 
$\mu_{0,\mathrm{LMC}}= 18.94 \;\mathrm{mag}$
with an internal error of $0.04\;\mathrm{mag}$; this modulus decreases by about $0.04$ mag if calibrators are
half-weighted. More recently \citet{kovacs00} (with a method based on double mode Cepheids), find 
$\mu_{0,\mathrm{LMC}}= 19.05 \pm 0.13 \;\mathrm{mag}$
and \citet{cioni_etal_00} have $\mu_{0,\mathrm{LMC}}= 18.99 \pm 0.03$ (formal) $\pm 0.08$ 
(systematic) mag. We retain the following estimation:
$$
    \mu_{\mathrm{SMC}} = 18.99 \pm 0.10 \;\mathrm{mag}
$$
 The SMC distance modulus only represents an average distance.
\citet{crowl_etal_01} have evaluated the depth of the SMC along the line-of-sight by
a study of populous clusters. They derived a depth between $\sim 6$ kpc and
$\sim 12$ kpc; these values lead to magnitude differences of $0.2$ and $0.4$ mag
respectively. Previous studies, see for instance \citet{gardiner_etal_91}, show similar results
with a line-of-sight SMC depth ranging between $\sim 4-7$ kpc and $\sim 15$ kpc
strongly depending on the location in the SMC.

 We have chosen to model the SMC depth with a gaussian distribution of 
distances around
$\mu_{\mathrm{SMC}}$ with a standard deviation: 
$$
\sigma_{\mathrm{SMC}}^{\mathrm{depth}} = 0.05 \;\mathrm{mag}
$$
which represents a total depth of $\sim 8$ kpc (about $\sim 0.3$ mag).
\subsection{\label{red_abs}Reddening and absorption}

 We have to distinguish: \textbf{foreground} reddening $\mathrm{E(B-V)_{MW}}$
(due to material in Milky Way) and \textbf{internal} reddening $\mathrm{E(B-V)_{i}}$ with an origin into 
the Cloud itself. These quantities are expected 
to change along the line-of-sight. Here we model the total reddening as 
$\mathrm{E(B-V)_{MW+i} = E(B-V)_{MW} + E(B-V)_{i}}$ taking into 
account its
non-uniformity. 
From the literature, we derive estimations for the mean value
 and the dispersion of $\mathrm{E(B-V)_{MW+i}}$,
object-to-object variations can then be simulated. 

We now discuss reddening determinations
for SMC and LMC.
 From a study of spectral properties of galactic  nuclei  behind the Magellanic
Clouds, \citet{dutra_etal_01} have evaluated the foreground and 
background reddenings for both Clouds.
For the LMC, they found an average spectroscopic 
reddening of $\mathrm{E(B-V)_{MW+i} = 0.12 \pm 0.10}$ mag. The 
uncertainties  essentially come from the determination 
of the stellar populations belonging to
background galaxies: in the case of LMC, when \citet{dutra_etal_01} consider 
only red population 
galaxies, they find $\mathrm{E(B-V)_{MW+i} = 0.15 \pm 0.11}$ mag, 
which gives an idea of the global
uncertainty on $\mathrm{E(B-V)}$, which 
 should be around $\sim 0.02 - 0.03$ mag 
(about $\sim 13 - 20 \%$). 
For the SMC \citet{dutra_etal_01} find $\mathrm{E(B-V)_{MW+i}
 = 0.05 \pm 0.05}$ mag.
\begin{center}
\rotatebox{-90}{\resizebox{8.0cm}{8.8cm}{\includegraphics{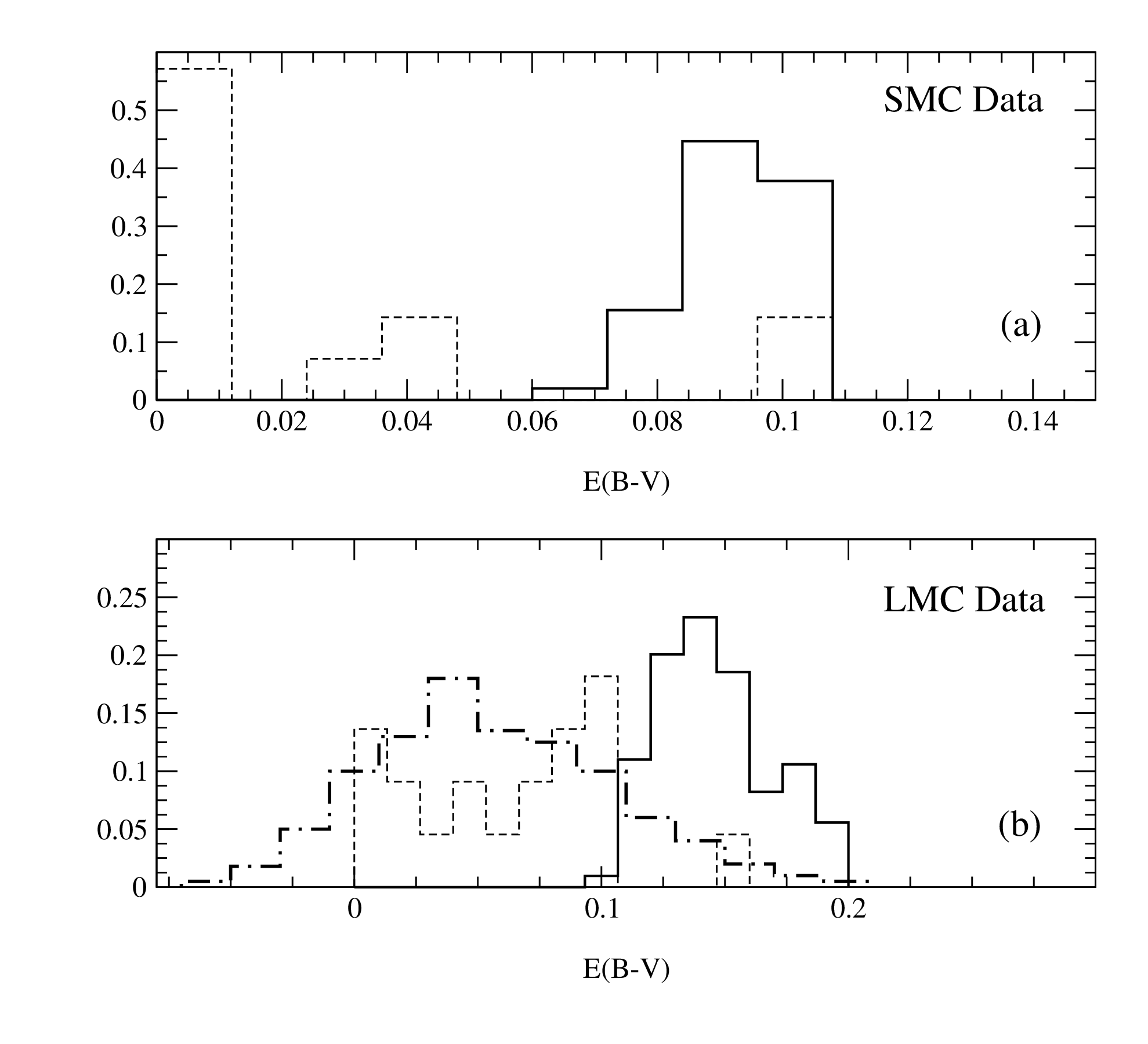}}}\\
\begin{figure}[h]
\caption[]{\label{distri_Reddening}Relative number of stars $N_{stars}/N_{tot}$ 
versus reddening from \citet{dutra_etal_01} (dashed curve), OGLE 2 experiment Cepheid
catalogue: see \citet{udalski_etal_99_a} and \citet{udalski_etal_99_b}, and data read in
Fig.~24 of  \citet{oestreicher_etal_95} (dot-dashed curve).  (a) Data for SMC, (b) 
data for LMC.}
\end{figure}
\end{center}
 The OGLE 2 project provides reddening for each Cepheid  star 
discovered in both Clouds.
OGLE values are: $\mathrm{E(B-V)_{MW+i}= 0.09 \pm 0.01}$ (SMC) and $\mathrm{E(B-V)_{MW+i}= 0.15 \pm 0.02}$
(LMC). In Fig.~\ref{distri_Reddening} we have displayed the histogram of $\mathrm{E(B-V)_{MW+i}}$ values from 
\citet{dutra_etal_01} and OGLE group. OGLE data have a better statistics with respectively 1333 (SMC) and 2049 (LMC)
objects, against 14 (SMC) and 22 (LMC) for \citet{dutra_etal_01}. Dutra et al's data are systematically less red, 
this could be inherent to their method: they observed objects \textbf{behind} Clouds and observations are 
easier through the more transparent regions of the clouds.

In addition, \citet{oestreicher_etal_95} have 
determined the reddening for 1503 LMC \textbf{foreground} 
stars with a $UBV$ photometry based method: 
$\mathrm{E(B-V)_{MW}= 0.06 \pm 0.02}$ mag, 
a quite low value because it is related to foreground stars. 
It shows a spread ($0.02$) similar to the OGLE 2 one.
\citet{oestreicher_etal_95} distribution is in very good agreement (see Fig.~\ref{distri_Reddening}b) with Dutra 
et al's one which tends to confirm that Dutra et al's result could be 
underestimated (Dutra et al's results are supposed to take account for foreground \textbf{and}
internal reddeing). Therefore in the case of LMC, we prefer to retain the OGLE average value
for purpose of consistency: 
$$
  <\mathrm{E(B-V)_{MW+i}^{LMC}}> = 0.15 \;\; \mathrm{mag}
$$
Fig.~\ref{distri_Reddening}b shows that the distribution shape is the same for \citet{dutra_etal_01} and
\citet{oestreicher_etal_95}, OGLE one being quite narrow which appears slightly underestimated, thus
we take a value similar to Dutra et al. one:
$$
  \mathrm{\sigma_{E(B-V)_{MW+i}}^{LMC}} = 0.08 \;\; \mathrm{mag}
$$
We take into account an additional 
uncertainty on $<\mathrm{E(B-V)_{MW+i}^{LMC}}>$ of about:
$$
  \delta_{\mathrm{E(B-V)}}^{\mathrm{LMC}} = 0.02 \;\; \mathrm{mag}
$$
The SMC case is more questionable (we only have two sets of data), 
we favor the OGLE values because they
are likely more suitable for performing
 simulations which synthesize OGLE data. Moreover OGLE data
have larger statistics. We adopt:
$$
  <\mathrm{E(B-V)_{MW+i}^{SMC}}> = 0.09 \;\; \mathrm{mag}
$$
with a crudely estimated uncertainty of about:
$$
  \delta_{\mathrm{E(B-V)}}^{\mathrm{SMC}} = 0.015 \;\; \mathrm{mag}
$$
In this case also, the OGLE standard deviation (see Fig.~\ref{distri_Reddening}a) seems to be low, 
therefore we adopt the \citet{dutra_etal_01} one:
$$
  \mathrm{\sigma_{E(B-V)_{MW+i}}^{SMC}} = 0.05 \;\; \mathrm{mag}
$$
 The absorption coefficient is taken from \cite{schlegel_etal_98}:
$$
     \mathrm{A_{V}} = 3.24 \times \mathrm{E(B-V)_{MW+i}}
$$
and is calculated for \textbf{each} object. 

\subsection{\label{bin_rate}Binary rate}

 Evaluating the average binary rate $<\beta>$ in objects as extended as the Magellanic Clouds is
not easy. Locally (i.e. within a particular area of the galaxy) this multiplicity rate depends -at least- on
two factors: (1) the star density and the kinematics of the objects which influence the encounter
probability; (2) the initial binary 
rate (relative number of binaries on the ZAMS). Within
 the Magellanic Clouds, the binary rate   likely varies over a wide  range 
and we only consider its spatial average value $<\beta>$. 

 \citet{ghez95} finds in solar neighbourhood that for main sequence stars and young stars the binary rate $<\beta>$
ranges between $0.10$ and $0.50$ (it peaks at $<\beta>= 0.50$). Therefore we tested the effects of
binarity for these two extreme values.

 In our population synthesis code, binaries are taken into account with a uniform probability for
the mass ratio $q=M_{2}/M_{1}$ (in the considered mass range).

\subsection{\label{imf_slope}IMF slope}

The IMF has been extensively discussed by many authors. Toward both Galactic poles
and within a distance of 5.2 pc from the Sun, \citet{kroupa93} found a mass function:
$dN/dm \approx m^{-\alpha_{\mathrm{Salp}}}$ with $\alpha_{\mathrm{Salp}}\approx 2.7$ for stars more 
massive than 1 $M_{\sun}$. In the LMC, \citet{holtzman97} inferred -from HST observations- a value
consistent with the \citet{salpeter55} one: $\alpha_{\mathrm{Salp}}\approx 2.35$. At very
low metallicity, \citet{grillmain98} observed the Draco Dwarf spheroidal Galaxy ($\mathrm{[Fe/H] \approx -2}$) 
with the HST. They concluded that the Salpeter IMF slope remains valid in the Magellanic Clouds and 
we have chosen: 
$$
\alpha_{\mathrm{Salp}}= 2.30 \pm 0.30
$$
 However we must keep in mind that  some circularity in work exists when using  an IMF. As described by 
\citet{garcia_mermilliod_01} the IMF can be derived from the observed Present Day Mass Function (PDMF) 
using evolutionary tracks and 
their corresponding time scales which depend on the adopted value for the 
overshooting !

\subsection{Star Formation Rate}

 For a given mass, the Star Formation Rate (SFR) represents the number 
of stars ``created'' per unit of time. 
\citet{vallenari_etal_96_b} have studied three stellar fields 
of the LMC and have  found a   
time scale of about  $\sim 2 - \sim 4$ Gyr for the
``bulk of star formation''. We therefore make the reasonable
assumption that the SFR remained quite constant 
during the short galactic period relevant for  this work, i.e. for
the last $\sim 300$ Myr. 
The SFR involved here is an average value over each cloud.

\section{\label{results}Resulting overshooting amounts}

\subsection{Large Magellanic Cloud}

\begin{figure*}
   \centering
\rotatebox{-90}{\resizebox{12cm}{18cm}{\includegraphics{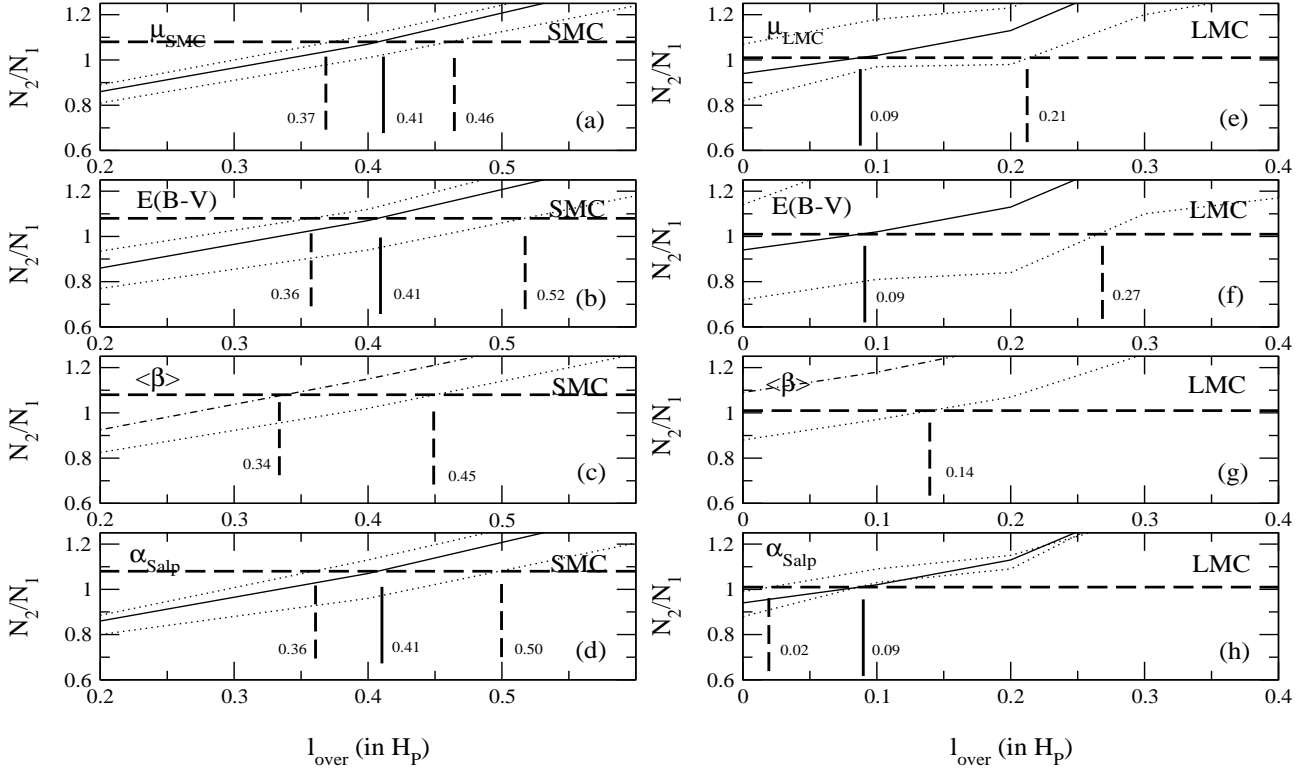}}}
      \caption{Overshooting determinations for SMC (panels a, b, c, and d) and LMC (panels e, f, g and h).
               The influence of distance modulus, reddening and IMF slope are considered for
               each cloud: continuous lines correspond to central values of these parameters discussed in
               Sect.~\ref{astro_in} and dashed lines to the associated error bars. Inferred values of 
               $l_{\mathrm{over}}$ (and uncertainties) are given on each figure. For (c) and (g) panels,
               dotted line is for $<\beta>=0.0$ and dot-dashed for$<\beta>=0.5$ (see text).}
         \label{det_ober_meth}
   \end{figure*}

 As a first step we choose the mean values for each astrophysical input
(discussed in Sect.~\ref{astro_in}), this yields for the LMC  the following overshooting:
$$
 l_{\mathrm{over}}= 0.09 \; \Hp
$$
which is a rather mild amount. We examine in Fig.~\ref{det_ober_meth}, how the
$l_{over}$-value is affected by the uncertainties on the astrophysical inputs:
\begin{itemize}
 \item Changing the IMF slope $\alpha_{\mathrm{Salp}}$ in the range $2.0-2.6$ we obtain:
$$
      0.02 \lesssim \; l_{\mathrm{over}}^{\mathrm{Salp}} \;
 \lesssim 0.09 \; \Hp
$$
       which tends to minimize the overshooting.
 \item Next, a test with the average binary
       rate $<\beta>$ in the range $0.10 - 0.50$ 
leads to:
$$
      0.00 \lesssim \; l_{over}^{<\beta>} 
\; \lesssim 0.14 \; \Hp
$$
 \item A distance modulus value in the range $18.35 \lesssim \mu_{\mathrm{LMC}} \lesssim 18.65$
       enables us to derive: $l_{over}= 0.0 \; \Hp$
       (in fact for $\mu_{\mathrm{LMC}}= 18.35$
       all the values for simulated $N_{2}/N_{1}$ are 
       larger than the observed one) and 
       $l_{over}= 0.21 \; \Hp$ for $\mu_{\mathrm{LMC}}= 18.65$.
 \item An average reddening between $0.13$ and $0.17$ 
       leads respectively to $l_{over}= 0.0 \; \Hp$
       (in this case also all the values for simulated $N_{2}/N_{1}$
       are larger than the observed one) and 
       $l_{over}= 0.27 \; \Hp$.
\end{itemize}

We stress that uncertainties on distance modulus and 
reddening infer the largest uncertainties on the final
overshooting values. We retain for the LMC average chemical composition:
$$
   l_{\mathrm{over}}^{\mathrm{LMC}} = 0.10^{+0.17}_{-0.10} \; \Hp
$$
which indicates that  a mild overshooting amount around $\sim 0.1 - 0.2 \; \Hp$ 
is needed to model  LMC stars as found in the majority of determinations involving
solar chemical composition objects (see Sect.~\ref{intro}).
\subsection{Small Magellanic Cloud}

 For the SMC, using the mean value of each astrophysical inputs we obtain
(see also Fig.~\ref{det_ober_meth}):
$$
 l_{\mathrm{over}}= 0.41 \; \Hp
$$
\begin{itemize}
 \item If the IMF slope varies between extreme values 
($-2.0 \lesssim \alpha_{\mathrm{Salp}} 
\lesssim -2.6$), the overshooting varies within the following boundaries:
$$
      0.36 \lesssim \; 
           l_{\mathrm{over}}^{\mathrm{IMF}} \; 
           \lesssim 0.50 \; \Hp 
$$
       similarly, an average binary rate ranging between 
       $0.10$ and $0.50$ leads to:
$$
      0.34 \lesssim \;
           l_{\mathrm{over}}^{<\beta>} \; 
           \lesssim 0.45 \; \Hp
$$
 \item The uncertainty on SMC distance modulus ($18.89 \lesssim \mu_{\mathrm{SMC}} 
       \lesssim 19.09$) leads to:
$$
      0.37 \lesssim \;
           l_{\mathrm{over}}^{\mu} \;
           \lesssim 0.46 \; \Hp
$$
 \item Similarly if one considers the uncertainty on the average reddening 
       ($<\mathrm{E(B-V)}>$ ranging between $0.075$ and $0.105$), the overshooting 
       amount shows a high sensitivity to reddening:
$$
      0.36 \lesssim \;
           l_{\mathrm{over}}^{\mathrm{E(B-V)}} \;
           \lesssim 0.52 \; \Hp
$$
\end{itemize}
Again, uncertainties on distance modulus and reddening are the largest.
We retain for the SMC:
$$
      l_{\mathrm{over}}^{\mathrm{SMC}} = 0.40^{+0.12}_{-0.04} \; \Hp
$$
Whatever the simulation is, statistical errors are of the order of 
$0.01 \; \Hp$ which can be safely neglected. In the SMC case, the required
overshooting appears to be much larger than for LMC stars and for solar
composition stars.

\section{\label{discu}Discussion}

\subsection{An upper limit with Roxburgh's criterion}

 Roxburgh's criterion \citep{roxburgh89} is a very general constraint on the size 
of the convective core. It is written as an integral formulation over the stellar
core radius:
\eqn{\label{rox_crit}
    \int_{r=0}^{r=R_{core}} (L_{rad}-L_{nuc})\frac{1}{T^{2}}  \frac{dT}{dr} dr 
  = \int_{r=0}^{r=R_{core}} \frac{\Phi}{T} 4 \pi r^{2} dr
    }
where $L_{rad}$ and $L_{nuc}$ are respectively the radiative energy flux
and the total energy flux (in J.s$^{-1}$) generated by nuclear processes, 
$r$ is the radius, $R_{core}$ is the core size including the ``overshooting'' region. 
$\Phi$ represents the viscous dissipation (in J.s$^{-1}$.m$^{-3}$). In the whole stellar 
convective core the turbulence  is supposed to be statistically stationary and the 
temperature gradient has to be almost adiabatic. 
 In Eq.~\ref{rox_crit} the integrand is positive when $r$ is lower than the Schwarzschild 
boundary where $L_{rad}=L_{nuc}$ and it becomes negative beyond. 

The viscous dissipation $\Phi$ is unknown but the integral 
constraint is satisfied for larger $R_{core}$ value when $\Phi=0$.  
Hence, neglecting the  dissipation by
setting $\Phi=0$ provides  the maximum possible extent of the convective core
which can be considered as the upper limit for overshooting. 
Evolutionary tracks have
been calculated, using Roxburgh's criterion, 
for a representative mass of 6 
$\mathrm{M}_{\odot}$ and SMC and LMC metallicities. The equivalent overshooting amount (EOA),
given in Table 1, is the time weighted average overshooting distance along the evolutionary
tracks, expressed in pressure scale height.

 In both cases (LMC and SMC), Roxburgh's criterion predicts a maximum value (i.e. neglecting
viscous dissipation) around $0.6 \; \Hp$ (see Tab.~\ref{rox_nodissip}) independent from $Z_{0}$.
Our determinations -i.e. 
$l_{\mathrm{over}}^{\mathrm{ SMC}}= 0.40^{+0.12}_{-0.06} 
\; \Hp$ 
and $l_{\mathrm{over}}^{\mathrm{ LMC}}= 0.10^{+0.17}_{-0.10} 
\; \Hp$- therefore are compatible with the theoretical upper limit given by the
Roxburgh's criterion.
\begin{table}
\caption[]{\label{rox_nodissip}Time weighted average overshooting distances for a 6 $\Msol$
main sequence model, derived with the ``Roxburgh's criterion'' neglecting dissipative
phenomenon ($\Phi=0$).}
\begin{tabular}{c|ccc}
\hline
Metallicity $Z_{0}$ & 0.004 (SMC)           & 0.008 (LMC)\\
Average EOA         & 0.6  $\mathrm{H_{p}}$ & 0.6    $\mathrm{H_{p}}$ \\
\hline
\end{tabular}
\end{table}

\subsection{\label{larotation}Influence of rotation}

 In addition to convection, rotation is an other important phenomenon inducing 
mixing through shear effects and other instabilities. For instance \citet{venn99}
finds surface abundance variations in SMC A supergiants that could be explained by
some kind of mixing related to rotation.

 Taking account of the rotational effect brings new important unknown features:
(1) the $\Omega$-value distribution and (2) the  $v \sin i$ distribution 
for the considered stellar population. Both features remain unconstrained by observational
studies.

In addition, stellar rotation involves many effects and 
physical processes that are non-trivial to include in modern evolutionary 
codes. \citet{talon97} show that 
(see their Fig.5) a rotating 1D-model with an initial surface velocity of 300 km.s$^{-1}$ 
leads to a main sequence track equivalent to an overshooting model using 
$l_{\mathrm{over}}= 0.2 \; \Hp$. Despite great theoretical efforts, a free parameter remains for
horizontal diffusivity in \citet{talon97} treatment of rotational mixing 
\citep[see][]{zahn_92}.

 Rotation changes the global shape of an evolutionary track, through two distinct effects:
(1) the material mixing inside the inner part of the star which brings more fuel into
the nuclear burning zones like overshooting, (2) the effective surface gravity modification leading to
color and magnitude changes (which depend on the angle between the line-of-sight and the rotational 
axis).
 In their Fig.6, \citet{maeder_meynet_01} show the influence of 
rotation on evolutionary tracks for low 
metallicity objects ($Z_{0}=0.004$). 
These tracks have been calculated taking into account:
(1) an ``average effect'' on surface, 
(2) the internal mixing. These tracks are very similar to those 
calculated with different overshooting amounts values.

An additional effect which needs to be discussed here  is 
the surface effect: modifications of colors and magnitudes of MS stars   due to  rotation 
(in absence of any mixing phenomenon) have been studied by \cite{maeder_peytremann_70} with uniformly rotating models.
Their Tab.~2 gives expected changes of $\mathrm{M}_{V}$ and $(\mathrm{B-V)}$ as a function of $\Omega$ (angular
velocity expressed in break-up velocity unit) and $v \sin i$ (this latter ranges from $0$ to $457$ km.s$^{-1}$, 
for a 5 $\Msol$ star). In this table standard deviation for  $\mathrm{M_{V}}$ and $\mathrm{(B-V)}$, 
are: $\sigma_{M_{V}}= 0.18$ mag and $\sigma_{(B-V)}= 0.01$ mag. Therefore the rotation effect has roughly
the same order of magnitude than present uncertainties on magnitudes and colors.
\begin{center}
\rotatebox{-90}{\resizebox{4.0cm}{8.8cm}{\includegraphics{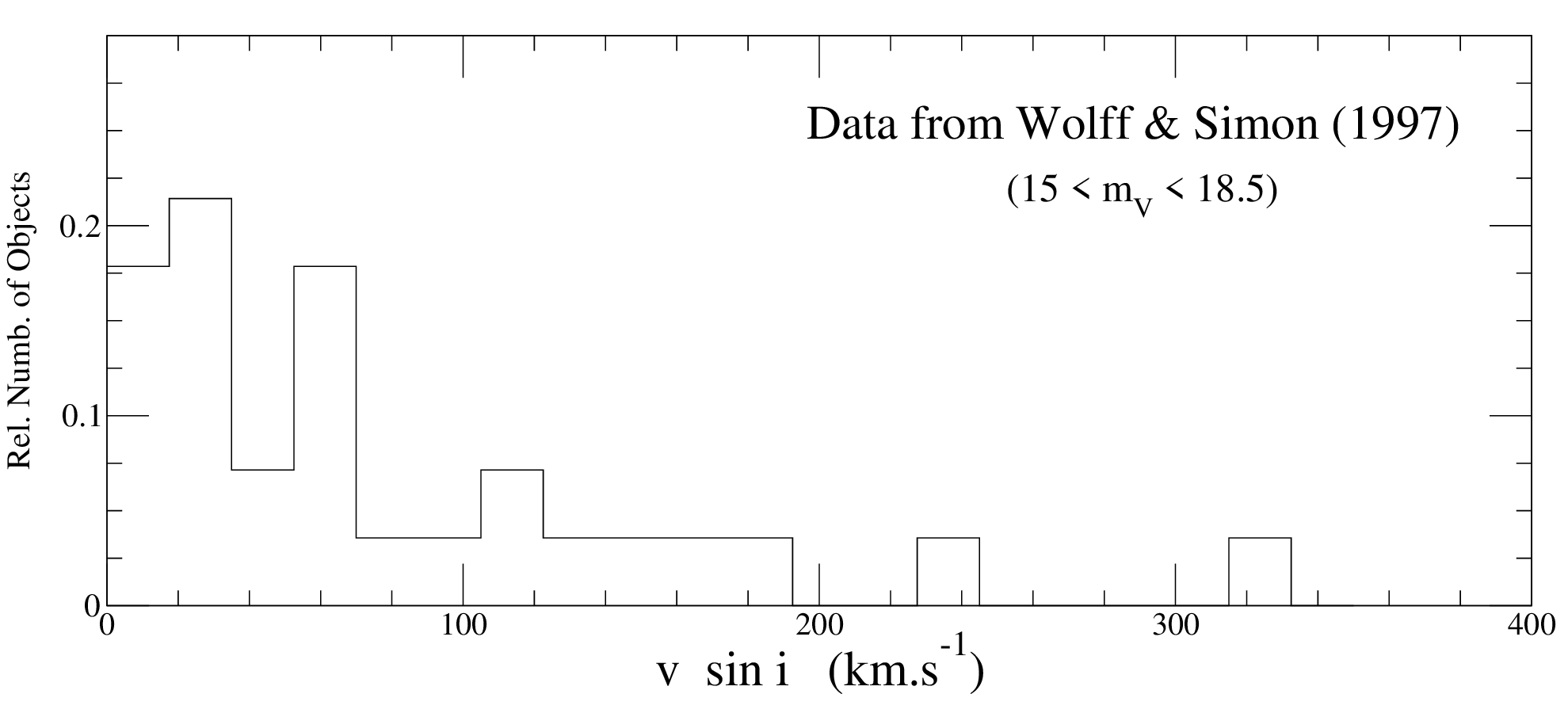}}}\\
\begin{figure}[h]
\caption[]{\label{distri_vsini}Relative number of stars ($N_{stars}/N_{tot}$ 
versus $V \times \sin i$ for \cite{wolff_and_Simon_97} objects with magnitudes in our studied range.}
\end{figure}
\end{center}
However, stars with $v \sin i$
greater than $\sim 200$ km.s$^{-1}$ are quite rare, as shown by the data of \citet{wolff_and_Simon_97} 
(see Fig.~\ref{distri_vsini}). Then keeping only data with $v \sin i \lesssim 200$ km.s$^{-1}$, leads to: 
$\sigma_{M_{V}}= 0.20$ mag and $\sigma_{(B-V)}= 0.001$ mag. The effect on absolute magnitude remains of 
the same order, whereas the effect on color becomes largely negligible. We conclude that our
results remain valid, even if the major mixing is due to rotation. In this case, the value of
$l_{\mathrm{over}}$ would change its meaning. Major contribution to $l_{\mathrm{over}}$
value  would represent a shear effect mixing.

\subsection{Influence of chemical composition gradient}

We have sofar assumed a uniform chemical composition.
The chemical composition may vary inside each Magellanic Cloud. The existence of an abundance gradient
in the Clouds is still debated and  spectroscopic measurements with
a statistics as large as the statistics of OGLE 2 data are not available. 
In their Tab.~4, \citet{luck_et_al98} give spectroscopic
determinations of $[\mathrm{Fe/H}]$ for 7 SMC Cepheids and 10 LMC Cepheids. 
For SMC data, the standard deviation is 
$\sigma^{\mathrm{SMC}}_{[\mathrm{Fe/H}]} \sim 0.07$ dex leading to negligible variations
for the heavy elements mass fraction $Z_{0}$. Therefore the SMC can be considered as chemically homogeneous
for our purpose. For LMC, \citet{luck_et_al98} find a standard deviation 
$\sigma^{\mathrm{LMC}}_{[\mathrm{Fe/H}]} \sim 0.10$ dex giving $0.007 \lesssim Z_{0} \lesssim 0.01$.
From evolutionary tracks of typical mass (6 $\Msol$) and an overshooting of $0.1 \;\mathrm{H_{P}}$,
changing $Z_{0}$ from $0.007$ to $0.01$ has a  negligible effect on magnitude and an effect of
$\sim 0.003$ mag  on color, which is largely lower than the photometric errors. We conclude that
-in the light of the present knowledge- the chemical composition gradient does not change our
results significantly.

\subsection{Comparison with other works} 

 From the investigation of young clusters in the  Magellanic Clouds, 
\citet{keller_etal_2001} did not find
any noticeable overshooting dependence with metallicity. They obtained for NGC 330 ($Z_{0} \sim 0.003$) 
$l_{\mathrm{over}}^{\mathrm{NGC 330}}= 0.34 \pm 0.10 \; \Hp$, which is compatible with
our determination for the SMC: 
 $l_{\mathrm{over}}^{\mathrm{SMC}}= 0.40^{+0.12}_{-0.04} \; \Hp$. For NGC 2004
($Z_{0} \sim 0.007$) \citet{keller_etal_2001} got 
$l_{\mathrm{over}}^{\mathrm{NGC 2004}}= 0.31 \pm 0.11 \; \Hp$; while for similar metallicity
we derived 
$l_{\mathrm{over}}^{\mathrm{LMC}}= 0.10^{+0.17}_{-0.10} \; \Hp$ which is also compatible
with Keller et al.'s result. One can note that masses involved in our
simulations
 (average mass of $\sim 7 - 8\;\Msol$ with a standard deviation of $4 \;
 \Msol$)
 are higher than the \citet{keller_etal_2001} one 
(\textbf{terminus} masses in the range $9 - 12 \; \Msol$ for the four clusters). 
\citet{keller_etal_2001} do not discuss the influence of the uncertainty on distance modulus of the
clusters and use $\mu_{\mathrm{LMC}}= 18.45$ mag and $\mu_{\mathrm{SMC}}= 18.85$ mag. 

 \citet{ribas_etal_00} derive overshooting amounts from evolutionary models of galactic binary systems. 
For SZ Cen ($Z_{0} \sim 0.007$) they find $0.1 \lesssim l_{\mathrm{over}} \lesssim 0.2 \; \Hp$
 which is close to our value for the LMC, but the mass of SZ Cen is $2.32 \; \Msol$ and
some mass effect
cannot be avoided, therefore any comparison with the present results  must be considered with care.
\begin{center}
\rotatebox{-90}{\resizebox{4.0cm}{8.8cm}{\includegraphics{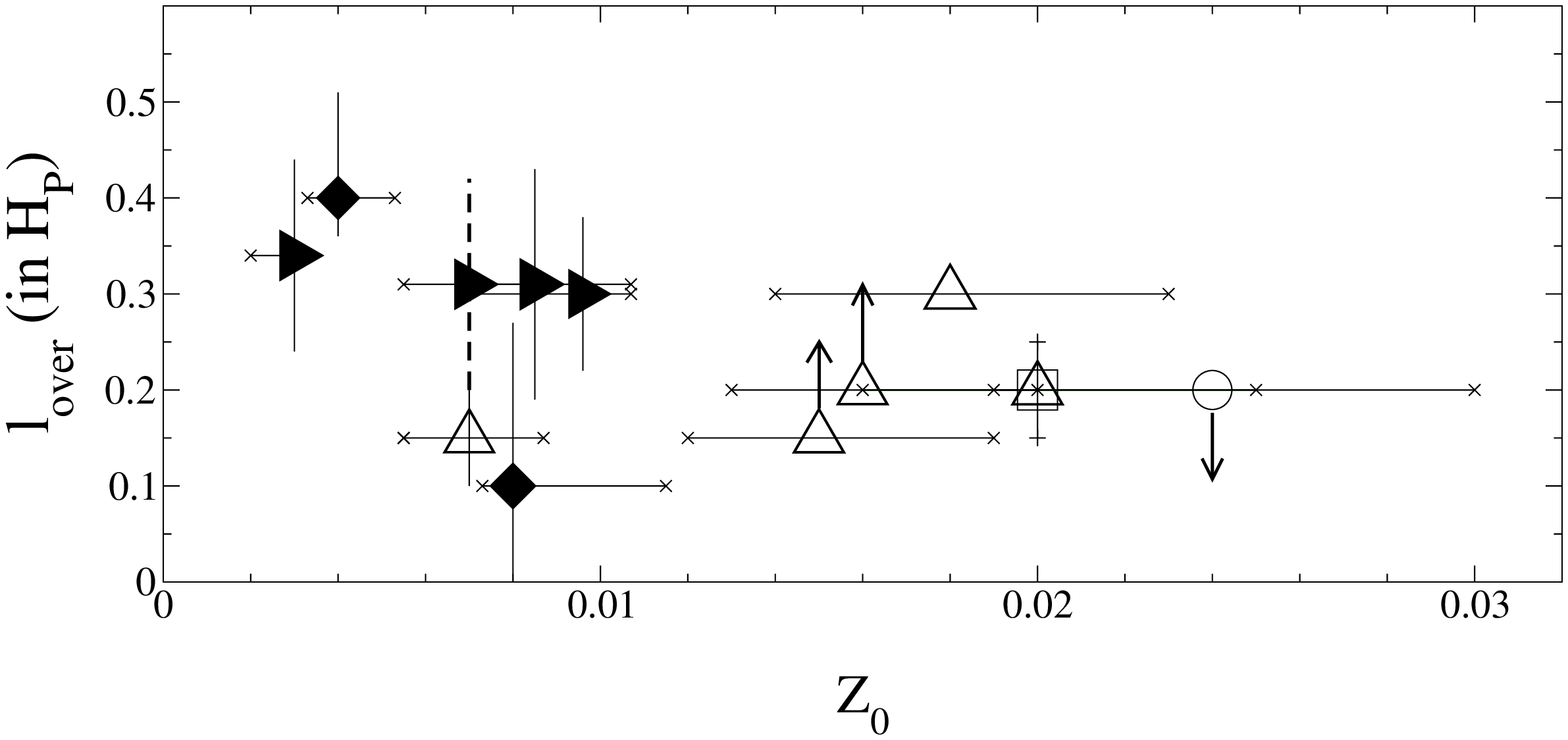}}}\\
\begin{figure}[h]
      \caption[]{\label{alpha_Z}Overshooting parameter $\alpha_{\mathrm{over}}$ versus metallicity
                 $Z_{0}$ from various sources. Open triangles represent results from
                 \citet{ribas_etal_00} for SZ Cen ($Z_{0} \sim 0.007$) error bars have been indicated,
                 arrows mean that the derived value is a minimum. The open square shows a result
                 from \citet{Kozhurina-Platais_etal_97} for the galactic cluster NGC 3680, error
                 bars are indicated. Filled triangles are determinations from \citet{keller_etal_2001}:
                 with continuous error bars amounts corresponding to the SMC cluster NGC 330, NGC 1818,
                 NGC 2100 and with dashed error bars result for the LMC cluster NGC 2004. Open circle: determination
                 in Hyades cluster from \citet{lebreton_etal_2001} (upper limit for overshooting).
                 Filled diamonds: SMC and LMC determinations performed in this work. Errors on $Z$ have been
                 evaluated assuming an internal error on $[\mathrm{Fe/H}]$ of 0.1 dex.}
\end{figure}
\end{center}
 In Fig.~\ref{alpha_Z} we summarize results from several authors. Despite the
small  number of points, a slight dependence of overshooting  with metallicity cannot be excluded.
However, at low and high metallicities, the considered  mass ranges are different and the errors remain
substantial, therefore a definite conclusion is not yet possible.

\section{\label{concl}Conclusion}

 In this paper we have estimated the overshooting distance from a 
 turbulent core for intermediate-mass main sequence stars. 
The result for SMC is $l_{\mathrm{over}}^{\mathrm{SMC}} = 0.40^{+0.12}_{-0.06} \;\mathrm{H_{P}}$,
and for the LMC $l_{\mathrm{over}}^{\mathrm{LMC}} = 0.10^{+0.17}_{-0.10} \; \mathrm{H_{P}}$.
The main contributions to errors are those brought by distance modulus
and reddening uncertainties. We have shown that chemical gradients within the  clouds and 
rotation surface effects of studied stars cannot significantly influence our
 results. 
Binary rate, IMF slope have no important effects as well. For SMC, despite different methods and
data, we find a result very similar to 
Keller et al.'s (2001) one for cluster
NGC 330.  The case of LMC is more questionable because of the rather large uncertainty 
on reddening.

 Fig.~\ref{alpha_Z} tends to indicate a sensitivity of overshooting to metallicity. However
a mass effect cannot be excluded, we can only stress that if such a dependence exists,
it should be an increase of overshooting with decreasing metallicity. However, the overshooting is
expected to increase with mass, unfortunately studied samples at solar metallicity have often lower masses than
those at low metallicities. Therefore further investigations are needed to disentangle these
effects. In any cases, if this dependence is confirmed the next challenge will be the physical explanation
of this metallicity-overshooting effect.

 Finally, the overshooting amounts derived in this work have a statistical meaning: they are
average values over time (in real star ``overshooting'' likely changes during the main
sequence) and over mass in the considered range. Moreover these amounts represent an extra-mixing
above the classical core generated either by  inertial penetration  of
convective bubbles  or
shear phenomena related to rotation. The real extent of the core likely
results from   a combination of both processes, indeed 
rotating models \citet{maeder_meynet_01}'s rotating models  still need  overshooting.

\begin{acknowledgements}
 We thank Jean-Paul Zahn and Ian Roxburgh for helpful discussions; we are also 
grateful to the OGLE group for providing their data and to Pierre Morel for writing the 
CESAM code. We thank the referee Dr. S.C. Keller for valuable remarks and suggestions.
\end{acknowledgements}
\bibliographystyle{apj}
\bibliography{apj-jour-perso,bibliographie_MASTER}

\appendix
\section{\label{appendix_noise}Photometric error simulations}

 As we selected the data using a criterion involving the photometric standard deviation of
magnitude measurements, we have to generate an artificial standard deviation for the theoretical
magnitude computed from evolutionary models. Moreover the general properties of the synthetic standard
deviation distribution must be similar to the OGLE 2 one.

\begin{figure}
   \centering
\rotatebox{-0}{\resizebox{9.15cm}{8.8cm}{\includegraphics{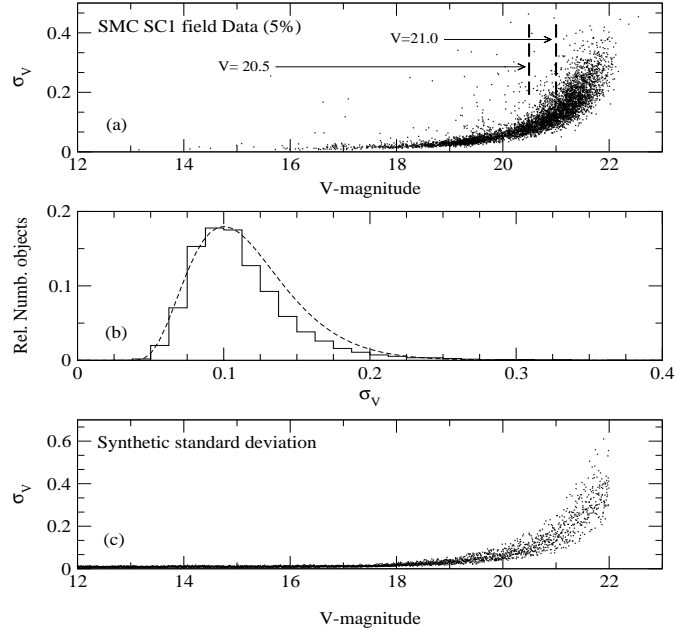}}}
      \caption{(a) Standard deviation $\sigma_{V}$ versus V-magnitude for objects
                   belonging to the SC1 field of the SMC. 
               (b) Histogram of the $\sigma_{V}$ values for magnitude V between 
                   20.5 and 21.0, the fit (dashed curve) is performed with a function
                   of a type given in the text. Differences between the fit and the histogram
                   are clearly insignificant for our purpose.
               (c) Synthetic $\sigma_{V}$ distribution generated with our algorithm.}
         \label{fig_noise}
   \end{figure}

 We describe here the scheme used to generate the pseudo-synthetic photometric
standard error distributions. The prefix ``pseudo'' means that we have extracted information
about the standard error distribution from the OGLE 2 data themselves (see Fig.~\ref{fig_noise}(a)).
For that purpose, we divide the relevant range of magnitudes into bins; in each bin, we construct
the histogram of standard deviation values (Fig.~\ref{fig_noise}(b)). This
histogram then is  fitted with a function of the form:

$$
 P(\sigma) = a \times (\sigma - \sigma_{min})^{4} \times e^{-b (\sigma - \sigma_{min})}
$$
where the constants $a$, $b$, $\sigma$, $\sigma_{min}$ are derived from the OGLE 2 data.
$P(\sigma)$ represents the probability for having the standard deviation $\sigma$.
The constants have been derived for each ``magnitude bin'', for each OGLE fields in
SMC and LMC. Then average values have been calculated over SMC and LMC.

 In our population synthesis code, for a given magnitude value $m$, a standard deviation
value $\sigma_{m}$ is  randomly  determined  following the probability law derived from OGLE.
After that, either the object is rejected (if the $\sigma_{m}$ value is too large) or the magnitude $m$ is 
changed into $m^{noisy}$, following a gaussian distribution having a standard deviation 
$\sigma_{m}$.

 Let us comment about differences between Fig.~\ref{fig_noise}(a) and 
Fig.~\ref{fig_noise}(c). Fig.~\ref{fig_noise}(a) contains the ``evolutionary information''
-i.e. more objects at high magnitudes- whereas Fig.~\ref{fig_noise}(c) does not contain this information,
objects have been uniformly distributed with respect to the magnitude. These facts explain the difference
between both figures.
\end{document}